# Light Scattering Measurements of KCl Particles as an Exoplanet Cloud Analog

Colin D. Hamill[1], Alexandria V. Johnson[1], and Peter Gao[2]
[1] Department of Earth, Atmospheric and Planetary Sciences, 550 Stadium Mall Drive, Purdue University, West Lafayette, IN 47907, USA; hamillc@purdue.edu
[2] Earth and Planets Laboratory, Carnegie Institution for Science, 5241 Broad Branch Road, NW, Washington, DC 20015, USA

## Abstract

Salt clouds are predicted to be common on warm exoplanets, but their optical properties are uncertain. The Exoplanet Cloud Ensemble Scattering System (ExCESS), a new apparatus to measure the scattering intensity and degree of linear polarization for an ensemble of particles, is introduced here and used to study the light scattering properties of KCl cloud analogs. ExCESS illuminates particles with a polarized laser beam (532 nm) and uses a photomultiplier tube detector to sweep the plane of illumination. Scattering measurements for KCl particles were collected for three size distributions representative of modeled clouds for the warm exoplanet GJ 1214b. Our measurements show that Lorenz–Mie calculations, commonly used to estimate the light scattering properties of assumedly spherical cloud particles, offer an inaccurate depiction of cubic and cuboid KCl particles. All of our measurements indicate that Lorenz–Mie scattering overestimates the backscattering intensity of our cloud analogs and incorrectly predicts the scattering at mid-phase angles (∼90°) and the preferential polarization state of KCl scattered light. Our results align with the general scattering properties of nonspherical particles and underscore the importance of further understanding the effects that such particles will have on radiative transfer models of exoplanet atmospheres and reflected light observations of exoplanets by the upcoming Nancy Grace Roman Space Telescope and Habitable Worlds Observatory.

*Unified Astronomy Thesaurus concepts:* Exoplanet atmospheres (487); Mini Neptunes (1063); Atmospheric clouds (2180); Laboratory astrophysics (2004); Interdisciplinary astronomy (804)

## 1. Introduction

Clouds and hazes are ubiquitous features of planetary atmospheres, and their presence has stark consequences for light scattering and atmospheric radiative balance (Hansen & Travis 1974; Marley et al. 2013; Gao et al. 2021). On exoplanets, clouds are typically inferred from muted atomic and molecular feature amplitudes in transmission spectra, as optically thick clouds prevent photons from probing deeper into the atmosphere (e.g., Knutson et al. 2014; Kreidberg et al. 2014; Crossfield 2015; Sing et al. 2016; Gibson et al. 2017; Wakeford et al. 2017; Bruno et al. 2018), impeding measurements of exoplanet atmospheric composition. Clouds have also been revealed through the observed westward shift of the optical phase curves of several hot Jupiters, such as Kepler-7b (Demory et al. 2013; Angerhausen et al. 2015; Muñoz & Isaak 2015; Parmentier et al. 2016). In addition, clouds can explain the thermal uniformity of hot Jupiter nightsides (Beatty et al. 2019; Keating et al. 2019; Gao & Powell 2021). However, despite these findings, current observations are only just beginning to understand the effects of cloud composition on exoplanets (e.g., Grant et al. 2023; Dyrek et al. 2024). A more rigorous understanding of how cloud particles scatter light across a variety of particle size distributions, shapes, and chemical compositions is therefore needed to interpret these observations more accurately, but nonspherical particles are often too complicated and time-consuming to be accurately modeled (e.g., Hovenier et al. 2003; Mishchenko & Travis 2003; Muñoz et al. 2004; Dabrowska et al. 2015). By directly measuring the light scattering properties of exoplanet cloud analogs through laboratory experiments, an improved understanding of cloud composition, atmospheric properties, and ultimately retrieved spectra and phase curves can be achieved.

To date, no experiments have been conducted to measure the scattering properties of exoplanet cloud analogs. However, a plethora of particle scattering measurements within the Earth and planetary science literature have shown why such data are valuable. As an example, measurements of volcanic ash and Martian dust analogs found that Mie scattering repeatedly under- or overestimates the measured scattering intensity, depending on the scattering angle, and resulted in largely negative degree of linear polarization (DOLP) values at most scattering angles, in contrast to the positive DOLP values that were measured (Muñoz et al. 2004; Dabrowska et al. 2015). Furthermore, models of the scattering effects of nonspherical particles, which are expected in many exoplanet atmospheres, show significant differences in scattering intensity and/or polarization compared to spherical particles (e.g., Mischenko et al. 1996; Moreno et al. 2007; Pinte et al. 2008; Dabrowska et al. 2013; Tazaki et al. 2016; Lodge et al. 2023). Since Lorenz–Mie calculations, which assume homogeneous spherical particles (Mie 1908), form the basis of how particle scattering is simulated in radiative transfer models (Seinfeld & Pandis 2016; Gao et al. 2021), it is important that we quantify the differences between spherical and nonspherical exoplanet cloud analogs. Once these differences are assessed, our laboratory measurements can then be used to better model the reflected-light properties of cloudy exoplanets.

In this paper, we present measured scattering intensities and DOLP for ensembles of potassium chloride (KCl) particles representative of those in exoplanet atmospheres. Thermochemical equilibrium condensation models predict that KCl can form clouds in warm (500 K < $T_{eq}$ < 1000 K) exoplanet and brown dwarf atmospheres at observable pressure levels ($\sim 10^{-2}$–1 bar; Lodders 1999; Morley et al. 2012). Most notably, KCl has been a proposed cloud species and atmospheric opacity source to explain the featureless infrared transmission spectrum (Berta et al. 2012; Fraine et al. 2013; Kreidberg et al. 2014; Kempton et al. 2023) of the mini-Neptune GJ 1214b (e.g., Kempton et al. 2011; Morley et al. 2012, 2013; Charnay et al. 2015; Gao & Benneke 2018; Ohno & Okuzumi 2018; Ohno et al. 2020). These studies have shown that KCl clouds can reproduce the observed transmission spectrum of GJ 1214b if the atmospheric metallicity is very high (1000x solar) and sedimentation efficiency is low (fsed ⩽ 0.1; Morley et al. 2013, 2015; Christie et al. 2022) and/or the atmospheric mixing is very high ($K_{ZZ} > 10^{10}$ cm$^2$ s$^{-1}$; Gao & Benneke 2018), signaling the presence of relatively small (submicron) particles.

Polarimetry is a valuable tool for understanding substellar atmospheres, motivating our measurement of the DOLP of KCl. Polarimetry has been used to characterize brown dwarf cloud properties (e.g., Ménard et al. 2002; Miles-Páez et al. 2013; Millar-Blanchaer et al. 2020), while for exoplanets, it can enhance planet–star contrast, making detections easier. Polarimetry also allows for more precise characterization of the atmospheres themselves, since the degree of polarization is linked to the composition, structure, and cloud coverage of an atmosphere (Stam et al. 2004; Stam 2008; Bott et al. 2018). A variety of studies have included polarization in their radiative transfer algorithms to predict the polarimetric signatures from the atmospheres of Earth-like or hot Jupiter planets (e.g., Seager et al. 2000; Stam et al. 2004; Stam & Hovenier 2005; Stam 2008; Bott et al. 2018; Groot et al. 2020; Gordon et al. 2023). Even though there has been no unambiguous detection of polarization from an exoplanet to date, the polarimetry of exoplanets is of strong interest for future missions and telescopes such as the Nancy Grace Roman Space Telescope, the Extremely Large Telescope, and the Habitable Worlds Observatory (Keller et al. 2010; Gaudi et al. 2020; Kasdin et al. 2020).

We use the Exoplanet Cloud Ensemble Scattering System (ExCESS) to measure the scattering intensities and DOLP of KCl particles at one discrete visible wavelength of 532 nm. ExCESS uses a single-wavelength laser to illuminate a small region of a continuous flow of particles and measures the scattering properties with respect to scattering angle along the scattering plane. We consider particle distributions with mean radii ranging from $\sim$0.6 to 1.2 $\mu$m and number densities from $\sim$160 to 1600 cm$^{-3}$, which were produced with either an atomizer containing a KCl and deionized water solution or a shaker flask containing a sample of dry KCl crystals. We then compare our measured quantities to those predicted by the Lorenz–Mie approximation for light scattering by spherical particles to evaluate how well it can reproduce observations of more realistic exoplanet cloud particle distributions. While the particle distribution parameter values were specifically chosen to match the results of atmospheric models of GJ 1214b (Charnay et al. 2015; Gao & Benneke 2018), our measurements of a diverse range of particle sizes are easily applicable to a wide variety of exoplanets and brown dwarfs that host KCl clouds.

In Section 2, we outline the light scattering theory and experimental techniques used to gather our measurements. In Section 3, we show our measured scattering intensities and DOLP curves for three size distributions of KCl and compare our measurements to Lorenz–Mie calculations. The implications and conclusions of our results are presented in Section 4.

## 2. Methods

### 2.1. Light Scattering Theory

We use Stokes parameterization to describe the scattering and polarization properties of light incident upon atmospheric particles. The mathematical descriptions here are used to describe single-scattering and multiple-scattering events (Hansen & Travis 1974). Consider an incident beam to be represented by two orthogonal components,

$$E_l = a_l^{i(\omega t - kz - \varepsilon_l)} \text{ and } E_r = a_r^{i(\omega t - kz - \varepsilon_r)}, \quad (1)$$

where $a_l$ and $a_r$ are the parallel and perpendicular amplitudes relative to the scattering plane, $\omega$ is the frequency, $\varepsilon_l$ and $\varepsilon_r$ are the phases, $t$ is the time, $i = (-1)^{1/2}$, and $k$ is the wavenumber related to the wavelength of light through $k = 2\pi/\lambda$ (Hansen & Travis 1974). We can then define the polarization of an incident beam by its Stokes parameters,

$$\begin{pmatrix} I \\ Q \\ U \\ V \end{pmatrix} = \begin{pmatrix} \langle a_l^2 + a_r^2 \rangle \\ \langle a_l^2 - a_r^2 \rangle \\ 2\langle a_l a_r \cos \delta \rangle \\ 2\langle a_l a_r \sin \delta \rangle \end{pmatrix}, \quad (2)$$

where the angled brackets indicate time averages, $\delta = \varepsilon_l - \varepsilon_r$, $I$ is the total intensity of the scattered light, $Q$ is the difference between parallel (0°) and perpendicularly (90°) polarized light relative to the scattering plane, $U$ is the difference between linearly polarized light at 45° and 135° relative to the scattering plane, and $V$ represents the difference between right-handed and left-handed circularly polarized light (Hansen & Travis 1974). We then define the DOLP as

$$-\frac{Q}{I} = -\frac{\langle a_l^2 - a_r^2 \rangle}{\langle a_l^2 + a_r^2 \rangle} = \frac{I_v - I_h}{I_v + I_h}, \quad (3)$$

where $I_v$ and $I_h$ are the vertically and horizontally polarized components of the scattered light with respect to the scattering plane viewed by the photomultiplier tube (PMT) detector.

While this description suffices for understanding the total intensity and DOLP for the scattered light measured in this paper, a more thorough description of particle scattering is often given with a 4 × 4 Mueller matrix. The Mueller matrix fully describes the relation between the incident and scattered Stokes parameters and consists of 16 elements that depend on the size, shape, and refractive indices of the scattering particles (van de Hulst 1957):

$$\Phi_{\text{det}}(\lambda, \theta) = \frac{\lambda^2}{4\pi^2 D^2} \begin{bmatrix} F_{11} & F_{12} & F_{13} & F_{14} \\ F_{21} & F_{22} & F_{23} & F_{24} \\ F_{31} & F_{32} & F_{33} & F_{34} \\ F_{41} & F_{42} & F_{43} & F_{44} \end{bmatrix} \Phi_0(\lambda, \theta), \quad (4)$$

where $\Phi_{\text{det}}$ and $\Phi_0$ are the Stokes vectors of the scattered light at each discrete scattering angle ($\theta$) at a given wavelength ($\lambda$) for the scattered and incident light, respectively. For the scattering angles used in this study, $\theta = 0°$ represents forward scattering and $\theta = 180°$ represents backscattering of the incident laser beam. The 16 matrix components, $F_{ij}$, completely describe the intensity and polarization properties of the scattered light. The distance from the detector to the scatterer is represented by $D$. When particles within the scattering region are asymmetric and randomly oriented, as we assume in this study, there are only 10 independent terms of the Mueller matrix (van de Hulst 1957):

$$F = \begin{bmatrix} F_{11} & F_{12} & F_{13} & F_{14} \\ F_{12} & F_{22} & F_{23} & F_{24} \\ -F_{13} & -F_{23} & F_{33} & F_{34} \\ F_{14} & F_{24} & -F_{34} & F_{44} \end{bmatrix}, \quad (5)$$

where $F$ is the full scattering matrix. By using Mueller matrix parameters instead of Stokes parameters, we may now also define our DOLP as

$$\text{DOLP} = \frac{I_v - I_h}{I_v + I_h} = -\frac{F_{12}}{F_{11}}, \quad (6)$$

where $F_{11}$ is proportional to the flux of scattered light and $F_{12}$ is related to the linear polarization of scattered light.

Table 1
Particle Generation Details for the Small, Medium, and Large KCl Size Distributions, Including the Mean Particle Radii and Particle Number Densities

| Particle Generation Parameters | Small KCl | Medium KCl | Large KCl |
|---|---|---|---|
| Method | Wet | Dry | Dry |
| Flow rate (lpm) | 0.85 | 0.2 | 0.5 |
| Weight | 0.25 g KCl/ 50 mL $H_2O$ | 4.0 g | 4.0 g |
| Grinding time (minutes) | N/A | 15 | 5 |
| Mean particle radius ($\mu$m) | 0.6 | 0.6 | 1.2 |
| Number density (cm$^{-3}$) | 160 | 460 | 1600 |

### 2.2. Experimental Apparatus

#### 2.2.1. Aerosolization Methods

We have implemented two separate aerosolization techniques in order to maximize our observed particle size range and particle morphology. The first method is wet particle generation, in which we create a KCl–$H_2O$ solution of 0.25 g KCl/50 mL $H_2O$ (Table 1). While the resultant particle size distributions can be altered slightly by changing the concentration of KCl, we found that the solution concentration produced a size distribution and particle number density comparable to modeled KCl clouds (e.g., Charnay et al. 2015; Gao & Benneke 2018). This solution is aerosolized with a constant output atomizer (TSI 3076), and a $N_2$ carrier gas flows at a constant rate of 0.85 liters minute$^{-1}$. The wet particles are then flown through a large in-line dryer (38 cm long, 9 cm diameter) filled with indicator silica gel beads to remove water from the solution drops, and the resulting dry KCl particles are flown through the inlet tube to the scattering region. This aerosolization method and noted solution strength produces KCl particles with a mean radius of 0.6 $\mu$m (Figure 1) as measured with an optical particle sizer spectrometer (TSI Model 3330). The polystyrene latex (PSL) spheres used to calibrate the instrument (see Section 2.2.4) are also aerosolized with the wet-generation method.

The second aerosolization technique is a dry particle generation method capable of aerosolizing larger KCl particles (Table 1). In the dry-generation method, dry KCl particles are placed in a conical flask attached to a shaker arm. A small stir rod placed inside the flask helps to agitate and aerosolize the particles. A $N_2$ carrier gas is then used to fly the particles into a second conical flask, a mixing volume, before reaching the scattering region of the ensemble system. This mixing volume helps regulate the particle number densities during our observational period, as it restrains large influxes of particles from entering the scattering region at any given moment. The resultant particle size distribution produced by this method can be altered by grinding the KCl particles in a mortar and pestle for varying amounts of time, where a longer grind time results in finer particles. Flow rate also alters the aerosol size distributions. Lower flow rates (<0.25 lpm) can only aerosolize smaller particles and thereby produce lower overall particle concentrations, while higher flow rates (>0.25 lpm) can aerosolize larger particles and produce higher particle concentrations. Here we use a 5 minute grind for our large-particle dry-generation experiments coupled with a 0.5 lpm flow rate and a 15 minute grind for our medium-particle dry-generation experiment with a 0.2 lpm flow rate to produce a varied size distribution range in conjunction with the wet-generation method (Figure 1). All three aerosolization techniques use a mass flow controller (Alicat ML-10SLPM-D/5M) to set the flow rates of the $N_2$ carrier gas.

Our wet-generation method, which involves desiccating $H_2O$–KCl solution droplets, produces submicron cubic crystalline particles (Walker et al. 2004). Our dry-generation method produces semicubic particles with morphological irregularities up to ∼5 $\mu$m in radius. Both morphologies produced in the lab are relevant to exoplanet cloud particles. The wet-generation method best represents homogeneous nucleation from gas to solid in a warm atmosphere without effects of fragmentation and/or coagulation. On the other hand, the dry-generation method better simulates a turbulent atmosphere in which nucleated KCl particles undergo a significant degree of mechanical deformation via particle collisions. It has been concluded that high rates of vertical transport are needed to reproduce the flat transmission spectrum of GJ 1214b using KCl clouds, suggesting that a high degree of particle shape irregularity may be present (Charnay et al. 2015; Gao & Benneke 2018). While it would also be of interest to aerosolize porous KCl particles (Ohno et al. 2020), neither of our current aerosolization techniques can produce significant amounts of coagulated KCl particles.

#### 2.2.2. ExCESS

A diode laser (OBIS LS 532 nm) illuminates the particles as a Hamamatsu PMT (part #R1288AH-27) with collection optics measures the intensity of light scattered by the particles on the scattering plane (Figure 2). The front of the PMT assembly is 90 mm away from the scattering region and is mounted to a precision nanorotator (Thorlabs part #HDR50) centered around the scattering region. The PMT can measure light scattered from 20° to 169°. At angles less than ∼20°, the intensity of light is too high for the PMT to measure without damaging the instrument. At angles higher than 169°, the PMT assembly physically blocks the incoming laser light from reaching the scattering region. The entire system is contained within a dark box, and any nonscattered laser light is terminated with a beam dump. The laser is shielded from the PMT with an anodized metal divider to further reduce stray-light detection from the PMT. The incoming laser reflects off two fused silica broadband (400–750 nm) dielectric mirrors at 90° and travels through a 3 mm diameter pinhole before entering the scattering region (Figure 2). The beam diameters are $0.8 \pm 0.1$ mm for both laser wavelengths used here.

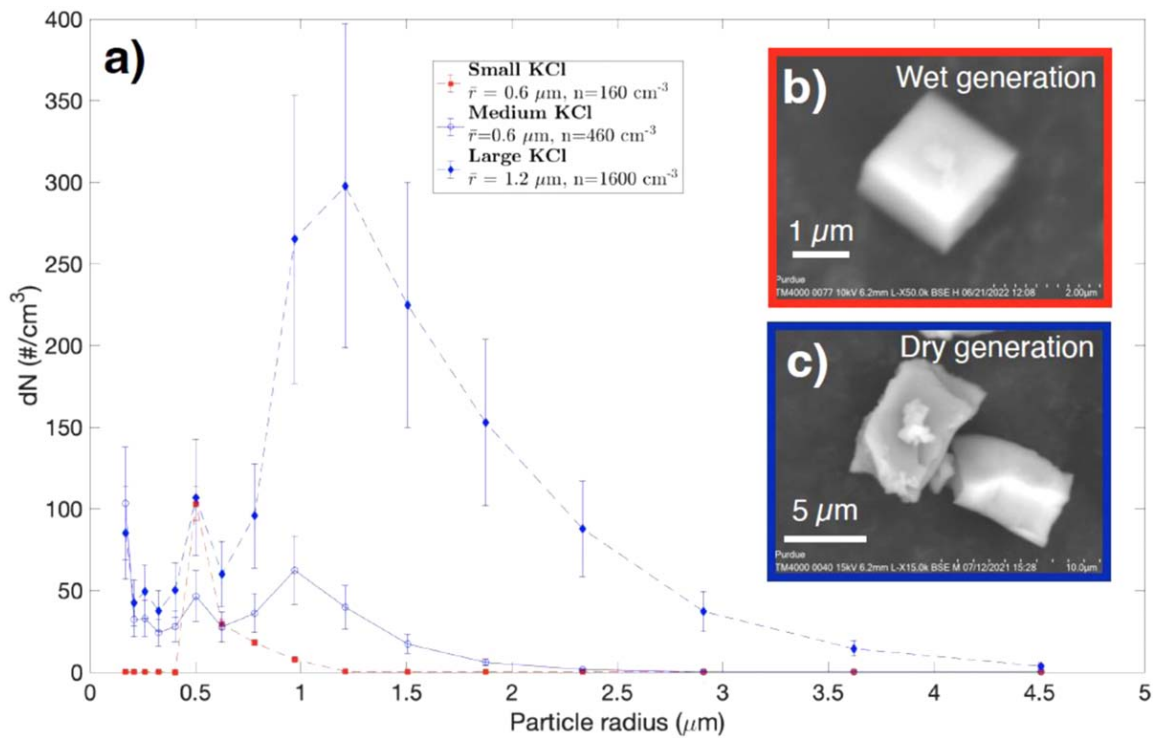

**Figure 1.** Size distributions and shapes of the three KCl cloud analogs used in this study. (a) Size distribution of the cloud analogs used here. The mean particle radii and total concentrations are provided in the legend. Our small KCl distribution is produced through wet generation (red), which creates a narrow size distribution. The medium (solid blue) and large (dashed blue) KCl distributions are produced through dry generation, which creates broader size distributions with larger particle number densities. (b) and (c) Scanning electron microscope images of small cubic KCl particles produced using wet generation and large cuboid and irregularly shaped KCl particles produced using dry generation, respectively.

The PMT is mounted to the nanorotator stage by Thorlabs cage plates (Thorlabs CP12) and an optical post. In front of the PMT, there is an achromatic doublet lens (Thorlabs MAP1030100-A) focused on the scattering region to collect light from the scattering particles and direct it upon the sensing area of the PMT face. In between the achromatic doublet and the PMT, there is an absorptive neutral density (ND) filter made of Schott glass void of polarization effects. This filter decreases the overall intensity of the light incident upon the PMT to ensure the light entering the PMT is kept under the recommended maximum intensity for operation. ND filters with optical densities of 0.2–0.6 were used for this study. The current from our PMT, which is the result of photons striking a photocathode to produce electrons that then strike a series of electron-multiplying dynodes, is converted into a voltage with a current-to-voltage converter (AAK LD15.1), where it is then read into our PC with a data acquisition module (National Instruments USB-6008). A program in LabVIEW records the PMT intensity data as a voltage at a rate of 120 Hz, capable of detecting incremental changes in intensity over a 30 s observation window per scattering angle.

The polarization of the incoming beam is set with a Glan–Thompson polarizer (Thorlabs GTH5M) mounted to the inlet of an electro-optic amplitude modulator (EOM; Thorlabs part #EO-AM-CR4, 400–600 nm). The lithium niobate ($LiNbO_3$) crystal within the EOM acts as a voltage-controlled wave plate that can alter the polarization state (i.e., vertical or horizontal) of the light depending on a modulated voltage supplied to the crystal. The Coherent OBIS LS/LX diode lasers used here are vertically polarized by a minimum polarization ratio of 100:1, so the Glan–Thompson polarizer is set to transmit vertically polarized light into the EOM. The EOM oscillates via a square wave between vertical and horizontal polarization at a frequency of 500 mHz, allowing us to gather the total intensity and the DOLP in a single run.

*2.2.3. Data Collection*

The nanorotator that holds the PMT is controlled with LabVIEW and steps in increments of 5° from 20° to 165° and increments of 1° from 165° to 169°. We change our angle increments at high scattering angles because backscattering effects are of particular interest in exoplanet atmospheric observations, so we are interested in gathering more precise data at these angles. Our PMT starts data collection at 169° and moves toward lower scattering angles. Before a scattering signal is measured, background noise is measured with the PMT at each angle. For background measurements, the laser is turned on, but there are no particles flowing into the scattering region. Background measurements and scattering measurements are taken for 35 s at each scattering angle. The first 5 s of each angle observation window are cut to account for the movement time of the PMT jogging between angles for both the background and scattering measurements. By averaging our observations over 30 s per scattering angle, we account for slight variations in particle output during our experiment.

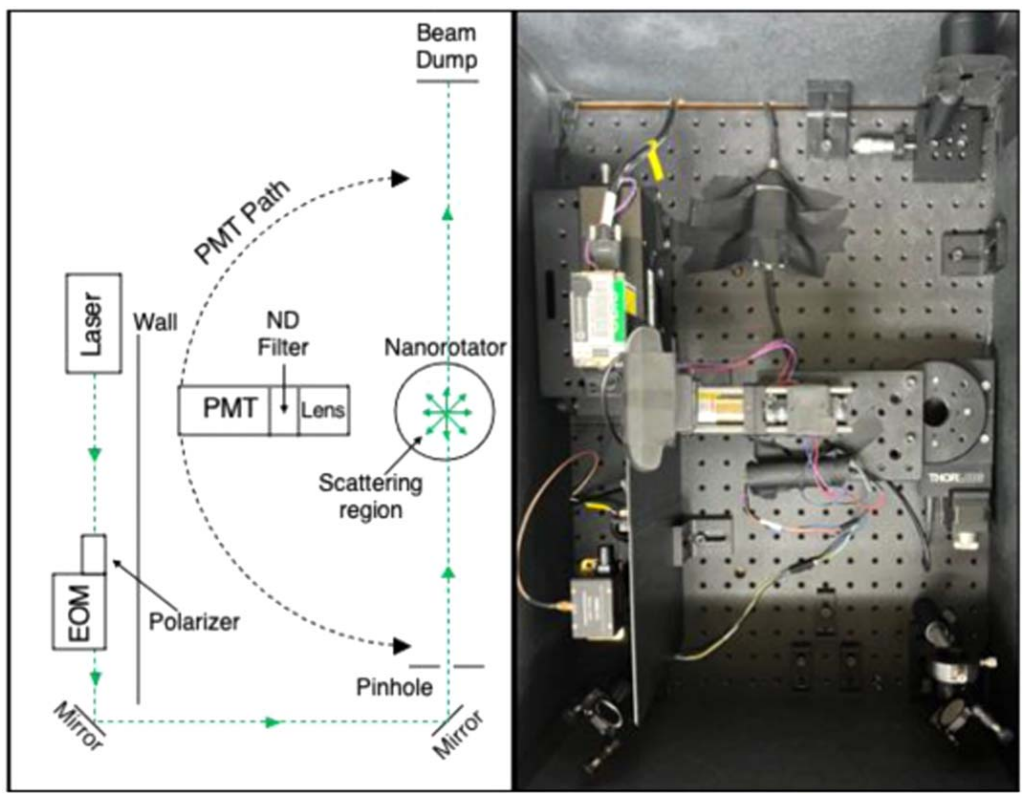

**Figure 2.** A top-down diagram and corresponding picture of ExCESS. Laser light is passed through an EOM that changes, and rapidly alternates, the polarization state of light before it is directed to the scattering region, where it impinges upon our cloud analog sample. An ND filter mitigates the intensity of scattered light entering the PMT detector. The PMT is mounted onto a stage that rotates around the scattering region where particles are flown in from above (into the page). ExCESS is a compact scattering system, contained within an 18' × 24' dark box, that collects measurements from 20° to 169°.

Directly after the background measurements are taken, we begin scattering data collection by aerosolizing particles and moving the PMT along the same path as the background measurement. Since intensity data are collected from the PMT at 120 Hz, there are roughly 3600 measurements per angle for both the background and scattering data regardless of the scattering angle. The background signal is subtracted from the scattering signal to account for any intensity measurements that are the result of stray light instead of particle-scattered light.

### 2.2.4. Calibration

To test the accuracy of ExCESS, we first measured the intensity and polarization properties of PSL spheres. PSLs are a useful and standard tool for ensuring instrument accuracy because they are spherical, have a known refractive index, and are manufactured to very precise particle sizes. Therefore, PSL particles scatter light in a way that can be characterized by Lorenz–Mie (hereafter referred to as Mie) scattering when the diameter of the PSL particles is comparable to the wavelength of the incident laser beam. Mie calculations describe the scattering of light by homogeneous spherical particles (Mie 1908). All Mie calculations in this paper are created with Philip Laven's MiePlot v4.6.

For our calibrations, we use 650 nm diameter PSLs. The PSLs are aerosolized with the same wet-generation method described above at a solution concentration of 30 drops of PSLs per 20 mL deionized water. The measured scattering properties of these PSLs are plotted in Figure 3 and compared to theoretical Mie curves. For the Mie calculations, we use refractive index values of $(1.60 + 5.72 \times 10^{-7}i)$ for 532 nm light (Zhang et al. 2020). The measured intensities are arbitrarily normalized to 1 at 30° since we cannot normalize our results to the peak intensity value at 0°. For comparison, the Mie curves are also normalized to 1 at 30°. Positive and negative DOLP curve values represent preferentially vertical and horizontal polarization, respectively.

The PSL data confirm that ExCESS can accurately measure the intensity and polarization of spherical particles. The intensity of PSL particles is a great match to Mie calculations at low and high scattering angles and slightly dimmer than Mie calculations at middle scattering angles (Figure 3). This effect is likely due to a slightly larger mean size of incoming PSL particles. PSLs sometimes coagulate to one another, resulting in a slightly more affected scattering pattern than their purported particle size (Lips et al. 1971). The dimmed intensity values at middle scattering angles may also be due to depolarization and extinction effects associated with flowing large concentrations (∼3000–4000 cm$^{-3}$) of PSLs into ExCESS's scattering region and the region directly encircling the scattering region. In practice, small monosize PSLs also scatter much less than our KCl particle distributions, meaning that the discrepancies seen in Figure 3 will be mitigated when measuring bright KCl scattering. Despite the differences in intensity between the measured PSLs and their corresponding Mie calculations, ExCESS still measures the intensity shapes and the forward- and backscattering intensities accurately. Furthermore, any extinction effects seen in the PSL measurements will be less prevalent in the KCl measurements, as the KCl measurements measure a much lower concentration of particles within the scattering region ($\leqq 1600$ cm$^{-3}$). The DOLP curve is slightly less polarized than the Mie calculations at low and high scattering angles (Figure 3). Depolarization is likely due to extinction effects as well.

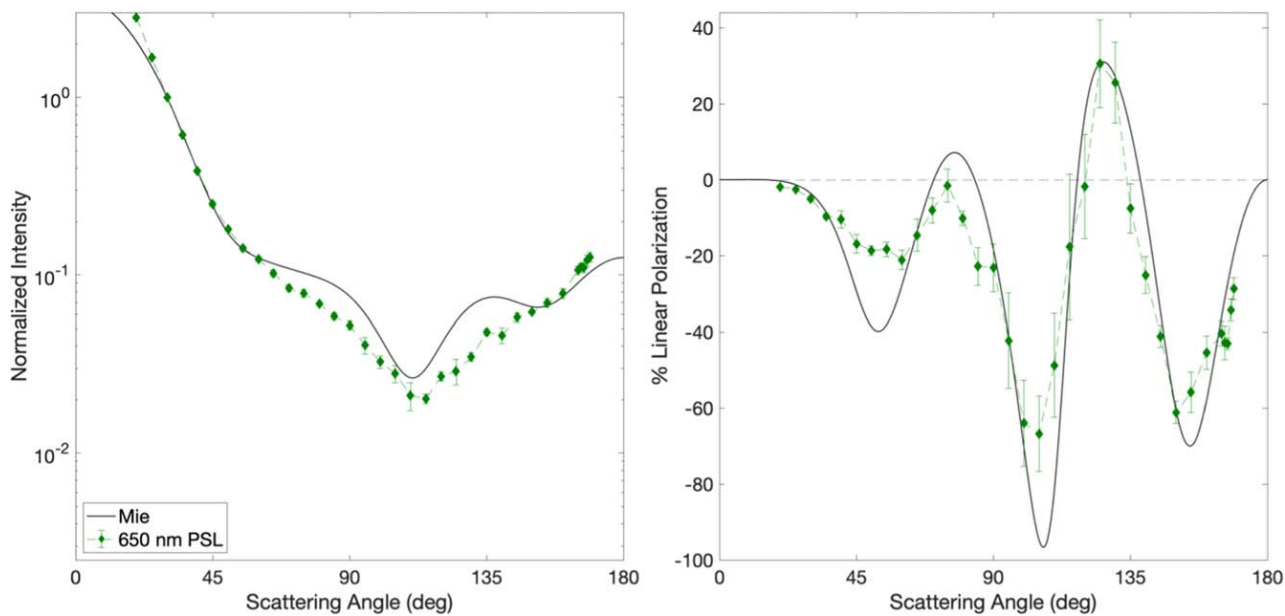

**Figure 3.** (Left) Measured PSL scattering intensity for 650 nm diameter PSLs using the 532 nm laser (green). (Right) DOLP for 650 nm diameter PSLs using the 532 nm laser (green), all compared to Mie calculations (black). Intensity values are normalized to unity at a scattering angle of 30°.

When aerosolizing our PSLs and KCl with the wet-generation method, we take great care in removing as much moisture as possible from our particles before entry into ExCESS. The silica gel beads within our system are regularly replaced, and particles are aerosolized with and flown within a dry $N_2$ carrier gas. This procedure ensures evaporation of water from our particles of interest (i.e., PSLs and KCl) and also mitigates the number of small (<100 nm) residual water droplets entering ExCESS. Our PSL measurements also produce the expected peaks and troughs of PSL scattering, indicating that small residual water droplets are not significantly altering our scattering measurements. Overall, our measured data match well to Mie calculations and show that our instrument is capable of collecting accurate scattering intensity and DOLP curves, despite the complicating factors arising from our large influx of particles.

## 3. Results

### 3.1. Measured Scattering Intensities and DOLP

We present scattering intensities and DOLP for three KCl size distributions measured from scattering angles of 20°–169° using the 532 nm laser in Figure 4. Every curve in Figure 4 represents an average of three to five individual scattering experiments; the error bars represent the standard error of the mean between individual runs, and in some cases the error bars are smaller than the plotted marker. Our calculated errors do not include those from other possible sources, such as slight misalignments of ExCESS or changes in incoming particle size distributions.

The small KCl size distribution, produced through wet generation, resulted in the dimmest intensities at mid- (45°–135°) and high (>135°) scattering angles. The medium KCl size distribution is slightly brighter than the small KCl size distribution. The intensities for the large KCl size distribution are brighter than the other size distributions and relatively flat across the mid- and high scattering angles (Figure 4(c)), with only minor features and small backscattering slopes at scattering angles greater than 160° (Figure 4(c)).

Overall, light scattered by KCl is preferentially vertically polarized for the majority of the scattering angles, though there is an overall tendency toward preferentially horizontal polarization states at scattering angles greater than ~145° (Figures 4(d)–(f)). The small size distribution produces the largest degrees of vertical polarization, with 52% peak vertical polarization (Figure 4(d)). The large KCl size distribution produces the smallest degrees of vertical polarization, with a peak of only 15% vertical polarization (Figure 4(f)). The peaks of the medium KCl size distribution have comparatively moderate degrees of polarization, with peak preferential vertical polarization of 18% (Figure 4(e)). The location of the DOLP peaks vary with size distribution. The small KCl size distribution shows an interesting bimodal degree of polarization at 532 nm that is not seen in the other measurements, with peaks at 85° and 125°. All the other DOLP curves exhibit a single peak in polarization between 85° and 115° (Figure 4). All DOLP curves tend to approach zero at low (<45°) scattering angles.

Every DOLP curve in Figure 4 shows an inversion angle, or the scattering angle at which the DOLP curve crosses from a positive to a negative value. Each inversion angle measured occurs between 140° and 160° (Figure 4). The small KCl size distribution measurements show the most negative DOLP values measured in this study, with a measured −20% linear polarization for 169°, the highest angle viewed with ExCESS (Figure 4(d)). The medium and large KCl size distribution measurements showed minimum values of −9% and −6%, respectively (Figure 4).

### 3.2. Comparison to Lorenz–Mie Calculations

While our KCl particles are not spherical, Mie calculations are used extensively to characterize cloud particle scattering (Seinfeld and Pandis 2016), particularly in exoplanet atmospheric models (Gao et al. 2021). Therefore, we offer a comparison of our measured data to Mie calculations to highlight the shortcomings of its use to describe the scattering and polarization properties of more realistically shaped cloud particles. The Mie calculations below use the measured size distributions for each KCl distribution shown in Figure 1. A real refractive index of 1.49 for KCl at 532 nm is used (Li 1976), and the KCl particles are assumed to be nonabsorbing (i.e., the imaginary refractive index is 0). The scattering intensities calculated from Mie theory are once again normalized to 1 at 30°.

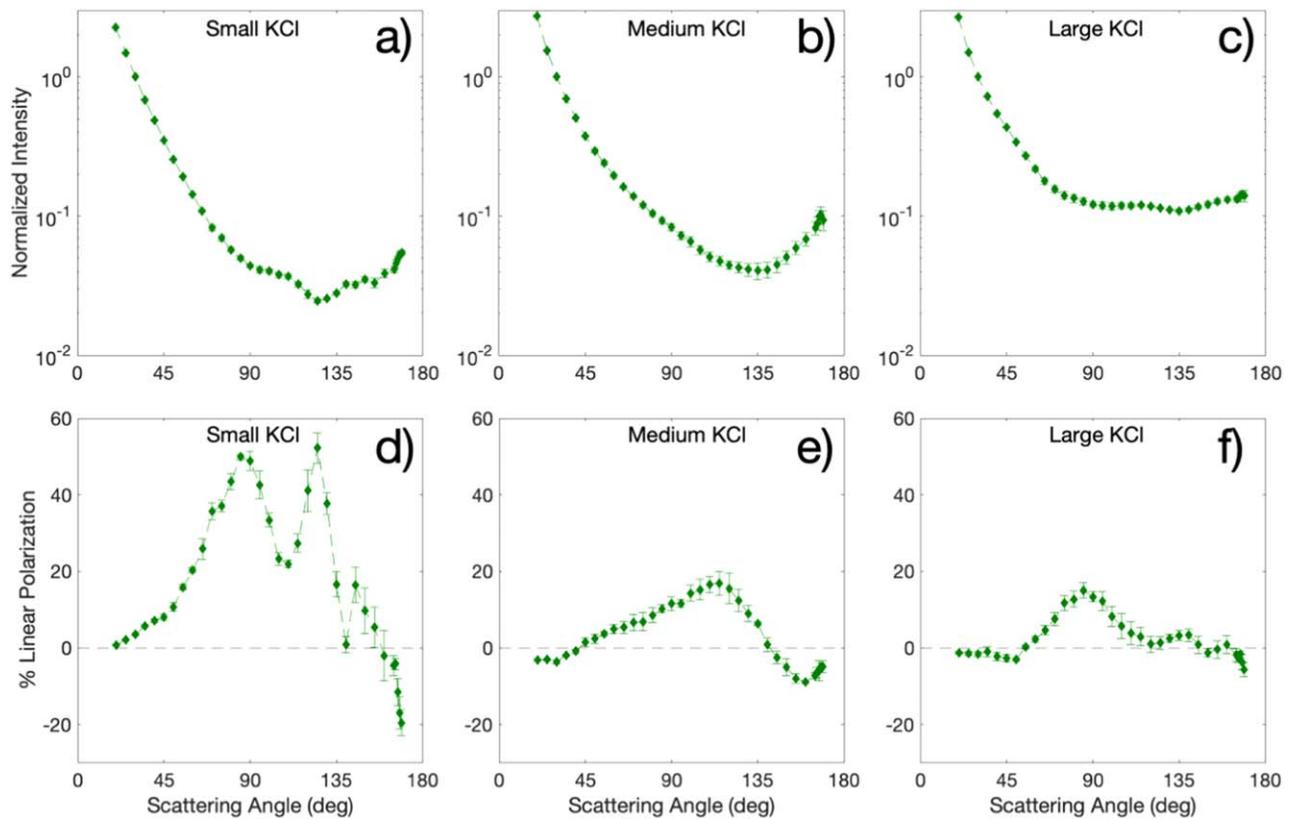

**Figure 4.** Measured KCl scattering intensities at 532 nm (green) for the (a) small, (b) medium, and (c) large KCl size distributions and DOLP curves for 532 nm for the (d) small, (e) medium, and (f) large KCl size distributions. Positive DOLP values represent preferentially vertically polarized light, and negative DOLP values represent preferentially horizontally polarized light. Intensity values are normalized to unity at a scattering angle of 30°.

The top row of Figure 5 compares our measured scattering intensities to Mie calculations, which show typical behavior for Mie scattering as a function of particle size: the small size distribution shows less overall variation in peak intensity as a function of scattering angle, while the large KCl size distribution shows enhanced forward scattering, reduced mid-angle scattering, and a steep backscattering slope. Our measured KCl scattering intensities, however, show almost the opposite effect. In Figure 5(a), the calculated Mie intensities overestimate the scattering intensity of our small KCl distribution at all angles greater than 40°. Conversely, the calculated intensities underestimate the scattering intensity of the large KCl distribution at mid-scattering angles. The Mie calculations for the medium size distribution, however, match well to the measured scattering intensities at mid-scattering angles. All three calculated intensities overestimate the backscattering intensity by up to an order of magnitude.

The bottom row of Figure 5 compares our measured DOLP curves to Mie calculations, which show predominantly negative values, indicating preferentially horizontally polarized light, with several peaks and troughs. Our measured KCl DOLP curves show that particles preferentially scatter vertically polarized light at scattering angles less than ∼135°, with little variation with scattering angle compared to Mie calculations.

### 4. Discussion and Conclusions

Our study provides new measured scattering intensities and DOLP curves for KCl particles at cloud-relevant size distributions for warm exoplanets and substantiates the general scattering characteristics of nonspherical particles measured for terrestrial and solar system applications (Bohren & Huffman 1998). Other laboratory measurements have shown similar scattering and/or polarizing characteristics for nonspherical particles with or without comparing their results to Mie calculations and have led to an increased level of precision in their respective fields (e.g., Muñoz et al. 2000, 2002, 2021; Volten et al. 2007; Renard et al. 2010; Muñoz & Hovenier 2011; Hadamcik et al. 2013; Escobar-Cerezo et al. 2018; Frattin et al. 2019; Levasseur-Regourd et al. 2019). Our results are in general agreement with these other studies while illustrating the impact of cubic and irregularly shaped cloud particles specifically for exoplanets and the utility in studying them in the laboratory. Exoplanet science stands to benefit greatly from an increased level of scattering precision due to the inherently low signal-to-noise and degeneracies present in reflected measurements. To further disentangle the scattering and polarizing effects of cubic versus irregularly shaped KCl particles, it may be necessary to aerosolize larger cubic particles using a higher concentration of wet-generated KCl or, conversely, aerosolize finer, irregularly shaped particles using our dry particle generation method.

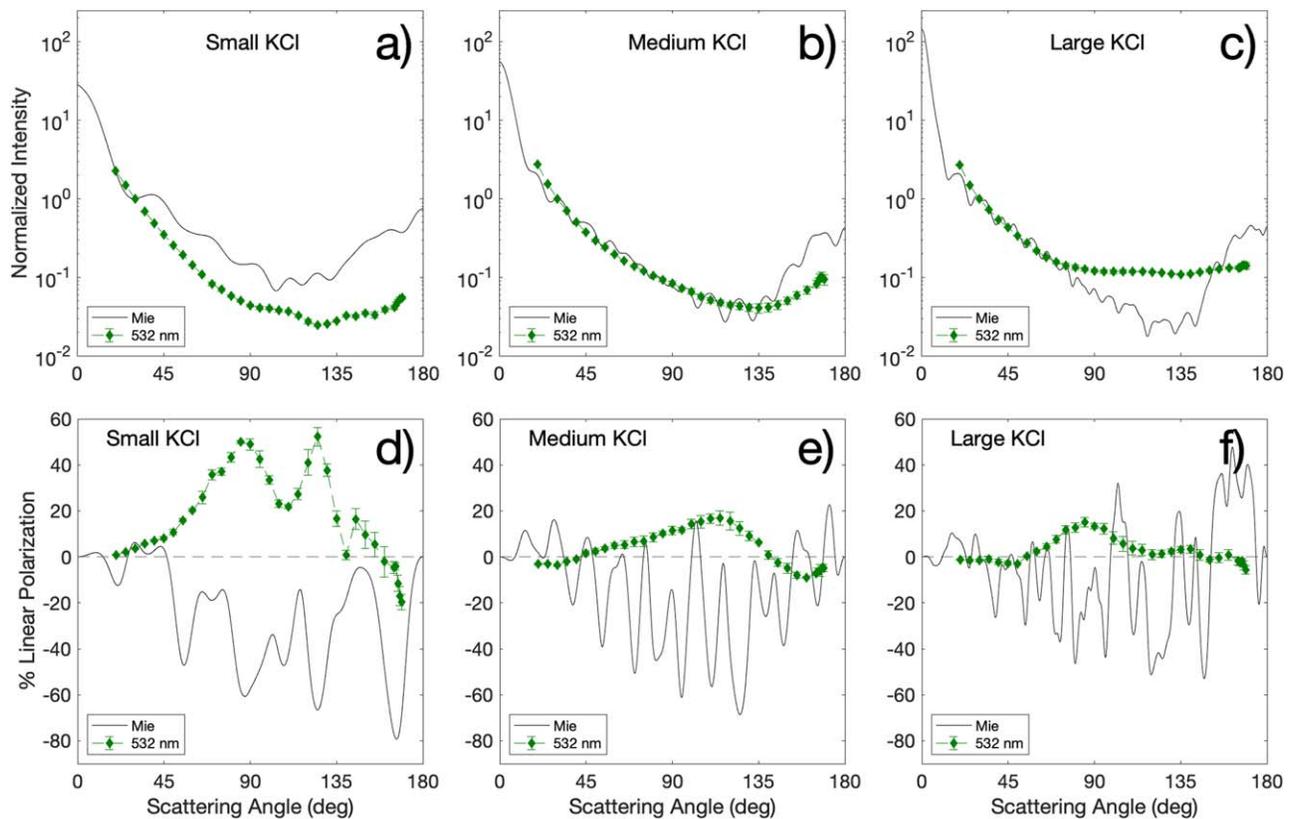

**Figure 5.** Measured KCl scattering intensities at 532 nm (green) for the (a) small, (b) medium, and (c) large KCl size distributions and DOLP curves at 532 nm for the (d) small, (e) medium, and (f) large KCl size distributions. Mie calculations for the measured size distributions and the refractive index of KCl are shown in black. Our Mie calculations use the size distributions of our three cloud analogs as measured in the laboratory (Figure 1(a)). Intensity values are normalized to unity at a scattering angle of 30°.

Our measurements with ExCESS demonstrate that Mie scattering is inadequate for modeling the intensity and DOLP curves of cubic and irregularly shaped KCl particles. These results suggest that radiative transfer models are insufficiently modeling the cloud particle scattering in exoplanet atmospheres. In particular, they are underestimating the scattering intensity for micron-sized particles or overestimating the scattering intensity of submicron-sized particles at mid-scattering angles. Current models are also severely overestimating the backscattering intensity of cloud particles across all size ranges. Both of these findings will significantly impact the observable features of exoplanets with KCl cloud cover. For example, secondary eclipse measurements of exoplanets with KCl clouds will likely be of lower intensity than expected from Mie calculations since backscattering effects are diminished with nonspherical particles. Conversely, our findings imply that nonspherical particles can lead to brightening of cloudy exoplanets when viewed in quadrature (mid-scattering angles), as in the case of direct imaging, when compared to Mie calculations. This suggests that future observations of warm and temperate exoplanets in reflected light by the Nancy Grace Roman Space Telescope and the Habitable World Observatory could yield greater signal-to-noise and allow for more precise measurements of Jupiter analogs and potentially habitable rocky exoplanets.

Our work highlights the need for more exoplanet-relevant particle scattering measurements, as there is a plethora of cloud species that will form as solid particles and thus show deviations from Mie scattering as a result of their irregular shapes (Gao et al. 2021). Further work is required to quantify the differences between modeled and measured light scattering for exoplanet clouds in order to make the most of current and future exoplanet observations across the parameter space of exoplanet temperatures. Employing techniques such as discrete dipole approximation could offer valuable insights into the scattering behavior of these cloud particles and aid in the interpretation of experimental data. Scattering measurements are also needed in the red and near-infrared range to extend our measured wavelength range and better interpret data from the James Webb Space Telescope (e.g., Gao et al. 2023; Kempton et al. 2023). As the radiative effects of clouds represent one of the biggest margins of uncertainty for understanding exoplanet climates, disentangling the effects of these clouds will require a momentous effort from observers, modelers, and experimentalists alike.

## Acknowledgments

We thank the anonymous reviewers for their helpful comments that improved the manuscript. We acknowledge support from the NASA Exoplanet Research Program (80NSSC23K0041). C.D.H. and A.V.J. acknowledge support from the Purdue University Research Scholars Grant. C.D.H. is thankful to Carmen Dameto de España, Justin Jacquot, and Xiaoli Shen for their valuable expertise and assistance in the laboratory. C.D.H. is thankful to Michelle Thompson for the use of the Hitachi TM4000Plus Scanning Electron Microscope. C.D.H. acknowledges the use of Philip Laven's program, MiePlot v4.6, for Mie calculations. A.V.J. is thankful for the early guidance of and discussions with Maria Zawadowicz and the efforts of Robert Washington and Kimberly Hernandez in initial setup and testing of individual components at Purdue and Brown University, respectively.

## ORCID iDs

Colin D. Hamill https://orcid.org/0000-0002-9464-8494
Alexandria V. Johnson https://orcid.org/0000-0002-6227-3835
Peter Gao https://orcid.org/0000-0002-8518-9601